\documentstyle[sprocl]{article}

\def\be{\begin{equation}}
\def\ee{\end{equation}}
\def\bea{\begin{eqnarray}}
\def\eea{\end{eqnarray}}
\newcommand{\beq}{\begin{equation}}
\newcommand{\eeq}{\end{equation}}
\newcommand{\Eq}[1]{equation~\ref{#1}}
\newcommand{\as}{\alpha_s}
\newcommand{\D}{\displaystyle}

\begin{document}

\begin{flushright}
hep-ph/9507236

FTUV 95/30

IFIC 95/32
\end{flushright}
\vspace{.5cm}

\title{LOW ENERGY YUKAWA INPUT PARAMETERS \\ FOR YUKAWA COUPLING UNIFICATION}

\author{G. RODRIGO}

\address{Dept. de F\'{\i}sica Te\`orica, Univ. de Val\`encia,\\
E-46100 Burjassot (Val\`encia), Spain}


\maketitle\abstracts{I make a short review of the most recent
determinations of the quark masses and run them to the $M_Z$
energy scale}


     GUT and SUSY theories predict some relations among the fermion
masses (or more properly among Yukawa couplings) at the
unification scale, $M_{GUT} \sim  10^{16} (GeV)$ .
For instance, in $SU(5)$ we have the usual lepton-bottom quarks
unification, $h_b=h_{\tau}$, $h_s=h_{\mu}$, $h_d=h_e$, or the
modified Georgi-Jarlskog relation, $h_b=h_{\tau}$, $h_s=h_{\mu}/3$,
$h_d=3 h_e$, while for $SO(10)$ typically we get unification of
the third family, $h_t=h_b=h_{\tau}=h_{\nu_{\tau}}$.
This together with the RGE's provide us a powerful tool
for predicting quark masses at low energies.

     In the following I will not talk about Unification but
rather try to make a short review of
the most recent determinations of the quark masses and to
do the running to $M_Z$, the mass of the $Z$ boson. Why $M_Z$?,
because for model building purposes it is a good idea to have
a reference scale, because extensions of the $SM$ appear above
$M_Z$ and because below $M_Z$ the strong coupling constant
$\as$ is really strong and then special care has to be taken
on the running and the matching in passing a heavy quark
threshold.

      Because of confinement to define what is the mass of a quark
is not an easy task. For leptons it is clear that the physical
mass is the pole of the propagator but for quarks we need
a precise theoretical framework and the masses appear more
like coupling constants. Dealing with quark masses we can
find the Euclidean mass, $M_E(p^2=-M^2)$, defined as the mass
renormalized at the Euclidean point $p^2=-M^2$, it is gauge
dependent but softly dependent on $\Lambda_{QCD}$.
However, it does not appear in the most recent works.
We can talk about the ``perturbative'' pole mass, $M(p^2=M^2)$,
perturbative because only order by order in perturbation
theory the pole of the propagator is well defined. It is
gauge and scheme independent but appears to be
ambiguous because of non-perturbative renormalons.
Finally $\bar{m}(\mu)$, the mass renormalized in the
Modified Minimal Substraction scheme or its corresponding
Yukawa coupling related to it through the
vev of the Higgs, $\bar{m}(\mu)=v(\mu) \bar{h}(\mu)$.

\begin{table}[t]
\begin{center}

\caption{Recent determinations of the light quark masses from
second order $\chi$PT, QCD Sum Rules and lattice.
\label{light0}}

\vspace{.3cm}

\begin{tabular}{|c|c|c|l|} \hline
J. Gasser  & $\chi$PT & $O(p^4)$
           & $\frac{\D m_d-m_u}{\D m_s-\hat{m}}
              \frac{\D 2\hat{m}}{\D m_s+\hat{m}}
             = 2.35 \times 10^{-3}$\\
H. Leutwyler & &
           & $m_s/\hat{m} = 25.7 \pm 2.6$ \\ \hline
Donoghue   & $\chi$PT & $O(p^4)$
           & $\frac{\D m_d-m_u}{\D m_s-\hat{m}}
              \frac{\D 2\hat{m}}{\D m_s+\hat{m}}
             = 2.11 \times 10^{-3}$\\
 et al.    & &
           & $m_s/\hat{m} = 31. $ \\ \hline
\end{tabular}

\vspace{.3cm}

\begin{tabular}{|c|c|c|l|c|} \hline
Bijnens    & FESR & NNLO
           & $(\bar{m}_u+\bar{m}_d)(1GeV)$
           & \\
Prades     & Laplace SR&
           & \qquad $= 12.0 \pm 2.5$
           & $\alpha_s(M_Z) = 0.117(5)$ \\
de Rafael  & (pseudo)&
           &
           & \\ \hline

Ioffe      & Isospin viol.& NNLO
           & $(\bar{m}_d-\bar{m}_u)(1GeV)$
           & $\Lambda = 150.$ \\
et al.     & in QCD SR &
           & \qquad $= 3.0 \pm 1.0$
           & \\ \hline
\end{tabular}

\vspace{.3cm}

\begin{tabular}{|c|c|c|c|c|} \hline
S. Narison & $\tau$-like SR& NNLO
           & $\bar{m}_s(1GeV) = 197(29)$
           & $\alpha_s(M_Z) = 0.118(6)$ \\
           & & NLO
           & $\bar{m}_s(1GeV) = 222(22)$
           & \\ \hline

M. Jamin   & QSSR & NNLO
           & $\bar{m}_s(1GeV) = 189(32)$
           & $\alpha_s(M_Z) = 0.118(6)$ \\
M. M\"unz  & (scalar) &
           &
           & \\ \hline

Chetyrkin  & QCD SR & NNLO
           & $\bar{m}_s(1GeV) = 171(15)$
           & $\alpha_s(M_Z) = 0.117(5)$ \\
et al.     & &
           &
           & \\ \hline \hline

Allton     & quenched  & NLO
                 & $\bar{m}_s(2GeV) = 128(18)$
                 & $\Lambda^5 = 240. \pm 90.$ \\
et al.     & Lattice &
           &
           & \\ \hline
\end{tabular}
\end{center}

\end{table}


     For the light quarks, up, down and strange, chiral perturbation
theory~\cite{GL,DON} provide us a powerful tool for determining
renormalization group invariant quark mass ratios. The absolute
values, usually the running mass at $1(GeV)$,
can be extracted from different QCD Sum Rules~\cite{BI,IO,NAs,JM,CH}
or for the strange quark mass from lattice~\cite{AL}.
For the heavy quarks, bottom and charm, we can deal either with
QCD Sum Rules~\cite{NAb,DOM,TY,NE},
lattice calculations~\cite{CRI,VG,DA,KH} or may be soon for
the bottom quark with jet
physics at LEP~\cite{RO95,FU}. For the top quark we have the
recent measurements from CDF an D\O\ at FERMILAB~\cite{top}
I will identify it with the pole mass.


\begin{table}[hbtp]
\begin{center}

\caption{Recent determinations of the heavy quark masses from
QCD Sum Rules, lattice and FERMILAB.
\label{heavy0}}

\vspace{.3cm}

\begin{tabular}{|c|c|c|c|c|} \hline
S. Narison & QSSR & NLO
           & $\bar{m}_b(M_b) = 4.23(4)$
           & $\alpha_s(M_Z) = 0.118(6)$ \\
           & $\Psi$, $\Upsilon$ &
           & $\bar{m}_c(M_c) = 1.23^{+(4)}_{-(5)}$
           & \\
           & &
           & $M_b = 4.62(2)$
           &\\
& &
           & $M_c = 1.42(3)$
           &\\ \hline

S. Narison & non-rel & NLO
           & $M_b^{NR} = 4.69^{+(3)}_{-(2)}$
           & $\alpha_s(M_Z) = 0.118(6)$ \\
           & Laplc.SR &
           & $M_c^{NR} = 1.45^{+(5)}_{-(4)}$
           & \\  \hline

Dominguez  & rel,non-rel & LO
           & $M_b = 4.70(7)$
           & $\Lambda^4 = 200-300$ \\
et al.     & Laplc.SR & $1/m_q^2$
           & $M_c = 1.46(7)$
           & $\Lambda^5 = 100-200$\\
           &  $J/\Psi$, $\Upsilon$ &
           & & \\  \hline

Titard     & $q\bar{q}$ &
           & $\bar{m}_b(\bar{m}_b) = 4.397^{+(18)}_{-(33)}$
           & $\alpha_s(M_Z) = 0.117(5)$ \\
Yndur\'ain & potential &
           & $\bar{m}_c(\bar{m}_c) = 1.306^{+(22)}_{-(35)}$
           & \\ \hline

M. Neubert & QCD SR & NLO
           & $M_b = 4.71(7)$
           & \\
           & & $1/m_q$
           & $M_c = 1.30(12)$
           & \\ \hline
\end{tabular}

\vspace{.3cm}

\begin{tabular}{|c|c|c|c|} \hline
Crisafulli & Lattice
           & $\bar{m}_b(\bar{m}_b) = 4.17(6)$
           & \\
et al.~\cite{CRI} & in B-meson
           &
           & \\  \hline

Crisafulli & Lattice
           & $\bar{m}_b(\bar{m}_b) = 4.15(7)$
           & \\
et al.~\cite{VG} & in B-meson
           &
           & \\  \hline

Davies     & NRQCD + leading
           & $M_b = 5.0(2)$
           & \\
et al.     & rel and Lattice
           & $\bar{m}_b(M_b) = 4.0(1)$
           & $\alpha^{(5)}_{\overline{MS}} = 0.115(2)$ \\
           & spacing, $b\bar{b}$
           & & \\  \hline

El-Khadra  & Fermilab action
           & $M_c = 1.5(2)$
           & \\
Mertens    & in quenched Lat
           &
           & \\ \hline
\end{tabular}

\vspace{.3cm}

\begin{tabular}{|c|l|c|} \hline
  CDF        & $M_t = 176. \pm 8.(stat) \pm 10.(sys)$ & {\it mean}\\ \hline
  D\O\       & $M_t = 199.^{+19.}_{-21.}(stat) \pm 22. (sys)$
& $M_t = 180. \pm 12.$ \\ \hline
\end{tabular}

\end{center}

\end{table}


     I have summarized in tables \ref{light0} and \ref{heavy0} all of
these recent determinations of the quark masses. Of course the final result
depends on the strong gauge coupling constant used in the analysis,
for this reason I quote it too. In the running I will take
$\as^{(5)}(M_Z) = 0.118 \pm 0.006$ for masses obtained from QCD Sum Rules
but for lattice masses I will run with the lattice~\cite{DA} result
$\as^{(5)}(M_Z) = 0.115 \pm 0.002$. This values are
consistent with almost all the references. For those that differ an
update is needed but this is beyond the goals of this paper. For instance,
S.~Narison~\cite{NAb} makes two different determinations
for the bottom and the charm quark masses.
In the first, and for the first time, he gets directly the running
mass avoiding then the renormalon ambiguities associated with the
pole mass. The second one, from non-relativistic Laplace Sum Rules,
is in fact an update of the work of Dominguez et al.~\cite{DOM}.

     The $O(\as^2)$ strong correction to the relation between the
perturbative pole mass and the running mass was calculated in~\cite{GB}
\beq
\frac{M}{\bar{m}(M)} = 1 + \frac{4}{3} \frac{\as(M)}{\pi}
+ K \left( \frac{\as(M)}{\pi} \right)^2 + O(\alpha_s^3(M)),
\label{fuerte}
\eeq
where $K_t \simeq 10.95$ for the top quark,
$K_b \simeq 12.4$ for the bottom and
$K_c \simeq 13.3$ for the charm.
As pointed out by S.~Narison~\cite{NAb} \Eq{fuerte} is consistent
with three loops running but for two loop running we can drop
the $O(\as^2)$ term. Recently, the electroweak correction to the
relation between the perturbative pole mass and the Yukawa coupling
has been calculated~\cite{HE}. However, this correction is
small, for instance for the top quark it is less than $0.5 \%$ in the $SM$
for a mass of the Higgs lower than $600 (GeV)$ and at most
$3. \%$ for $M_H \simeq 1 (TeV)$, and for consistency one has to
include it only if two loop electroweak running is done.

     Instead of expressing the solution of the QCD renormalization
group equations for the strong gauge coupling constant and the quark
masses in terms of $\Lambda_{QCD}$
we can solve the running as an expansion in the strong coupling
constant at one loop \cite{RO93}. At three loops

\bea
   \as(\mu) &=& \as^{(1)}(\mu) \left( 1 + c_1(\mu) \as^{(1)}(\mu)
                + c_2(\mu) (\as^{(1)}(\mu))^2 \right), \\
   \bar{m}(\mu) &=& \bar{m}^{(1)}(\mu) \left( 1 + d_1(\mu) \as^{(1)}(\mu)
                + d_2(\mu) (\as^{(1)}(\mu))^2 \right), \label{massrun}
\eea
where $\as^{(1)}(\mu)$ and $\bar{m}^{(1)}(\mu)$ are the one loop
solutions

\beq
   \as^{(1)}(\mu) = \frac{\as(\mu_0)}{1+\as(\mu_0) \beta_0 t}, \qquad
   \bar{m}^{(1)}(\mu) = \bar{m}(\mu_0) K(\mu)^{-2\gamma_0/\beta_0},
\eeq
with $t = 1/(4\pi) \log \mu^2/\mu_0^2$, $K(\mu)$ the ratio
$K(\mu) = \as(\mu_0)/\as^{(1)}(\mu)$, and

\bea
& &c_1(\mu) = - b_1 \log K(\mu), \nonumber \\
& &c_2(\mu) = b_1^2 \log K(\mu) \left[\log K(\mu) - 1 \right]
               - (b_1^2-b_2) \left[ 1 - K(\mu) \right], \nonumber \\
& &d_1(\mu) = - \frac{2 \gamma_0}{\beta_0} \left[
             (b_1-g_1)\left[1-K(\mu)\right] + b_1 \log K(\mu) \right], \\
& &d_2(\mu) = \frac{\gamma_0}{\beta_0^2} \biggl\{
             [ \beta_0(b_2-b_1^2) + 2\gamma_0(b_1-g_1)^2 ]
               \left[1-K(\mu)\right]^2  \nonumber\\
           &+& \beta_0 (g_2-b_1 g_1) \left[1-K^2(\mu)\right] \nonumber \\
           &+& \biggl[ 4\gamma_0 b_1 (b_1-g_1) \left[1-K(\mu)\right]
             - 2 \beta_0 b_1 g_1 + b_1^2(\beta_0+2\gamma_0) \log K(\mu)
             \biggr] \log K(\mu) \biggr\}, \nonumber
\eea
where

\bea
   b_1 &=& \frac{\beta_1}{4\pi \beta_0}, \qquad
   b_2 = \frac{\beta_2}{(4\pi)^2 \beta_0}, \qquad
   g_1 = \frac{\gamma_1}{\pi \gamma_0}, \qquad
   g_2 = \frac{\gamma_2}{\pi^2 \gamma_0},
\eea
are the ratios of the well known beta and gamma functions in
the $\overline{MS}$ scheme

\bea
     \beta_0&=&11 - \frac{2}{3} N_F, \qquad    \gamma_0=2, \nonumber \\
     \beta_1&=&102 - \frac{38}{3} N_F, \qquad
     \beta_2 = \frac{1}{2} \left( 2857 - \frac{5033}{9} N_F
                 + \frac{325}{27} N_F^2 \right), \nonumber \\
     \gamma_1&=&\frac{101}{12} - \frac{5}{18} N_F, \qquad
     \gamma_2 = \frac{1249}{32} - \frac{277 + 180 \zeta(3)}{108} N_F
                 - \frac{35}{648} N_F^2,
\label{beta}
\eea
and $\zeta(3) = 1.2020569 \dots$ is the Riemann zeta-function.
Our initial condition for the strong coupling constant will be
$\as(M_Z)$. Then we will run $\as$ from $M_Z$ to lower scales,
i.e. for instance $\mu_0 = M_Z$ or the upper threshold.
On the other side, for the masses we will run
from low to higher scales then we need the inverted version
of \Eq{massrun}

\beq
    \bar{m}(\mu_0) = \bar{m}(\mu) K(\mu)^{2\gamma_0/\beta_0}
\left( 1 - d_1(\mu) \as^{(1)}(\mu)
         + ( d_1^2(\mu)-d_2(\mu) ) (\as^{(1)}(\mu))^2 \right).
\eeq

     The beta and gamma functions depend on the
number of flavours $N_F$ therefore we have to decide where we have
five where we have four flavours. The trick is as was done by~\cite{BE}
and recently corrected by~\cite{LA} to built below
the heavy quark threshold an effective theory where the heavy
quark has been integrated out. Imposing agreement of both
theories, the full and the effective, at low energies
they wrote $\mu$ dependent matching conditions that
express the parameters of the effective theory, with $N-1$
quark flavours, as a perturbative expansion in terms of the
parameters of the full theory with $N$ flavours

\bea
   \as^{N-1}(\mu) &=& \as^N(\mu) \left[ 1
        + \frac{x}{6} \frac{\as^N(\mu)}{\pi}
        + \frac{1}{12}
\left(\frac{x^2}{3} + \frac{11x}{2} + \frac{11}{6} \right)
\left(\frac{\as^N(\mu)}{\pi}\right)^2 \right], \nonumber \\
   \bar{m}^N_l(\mu) &=& \bar{m}^{N-1}_l(\mu) \left[ 1
        - \frac{1}{12} \left( x^2 + \frac{5x}{3} + \frac{89}{36} \right)
\left(\frac{\as^{N-1}(\mu)}{\pi}\right)^2 \right],
\label{continuous}
\eea
with $x = \log \bar{m}^2(\mu)/\mu^2$,
where $\bar{m}(\mu)$ is the mass of the heavy quark we decouple at the
energy scale $\mu$ and $\bar{m}_l(\mu)$ are the masses of the light quarks.
This matching conditions make the strong coupling
constant and the mass of the light quarks discontinuous at the
thresholds. However, taking the matching in this way we ensure,
as pointed out explicitly in \cite{RO93}, that the final result
is independent of the particular matching point we choose for
passing the threshold. As it is independent, the easiest way to implement
a heavy quark decoupling is to take the threshold
as the running mass at the running mass scale, i.e.
$\mu_{th} = \bar{m}(\bar{m})$ or equivalently $x=0$,
then the discontinuity appears only at two loops matching.


\begin{table}[hbtp]
\begin{center}

\caption{Running at the NLO and NNLO of the top quark mass to $M_Z$,
$\as^{(5)}(M_Z) = 0.118 \pm 0.006$, $\as^{(6)}(M_Z) = 0.117 \pm 0.006$.
\label{heavytop}}

\vspace{.3cm}

\begin{tabular}{|l|l|l|l|} \hline
     & $\bar{m}_t(M_t)$ & $\bar{m}_t(\bar{m}_t)$ & $\bar{m}_t(M_Z)$ \\ \hline
NLO  & $172.\pm12.$ & $173.\pm12.$ & $182.\pm13.$ \\ \hline
NNLO & $170.\pm12.$ & $171.\pm12.$ & $180.\pm13.$ \\ \hline
\end{tabular}
\end{center}

\end{table}

\begin{table}[hbtp]
\begin{center}

\caption{Running at the NLO of the bottom and charm quarks masses to $M_Z$
and running masses at the running mass scale needed for thresholds.
For masses extracted from QCD SR
$\as^{(5)}(M_Z) = 0.118 \pm 0.006$, for lattice
$\as^{(5)}(M_Z) = 0.115 \pm 0.002$
\label{heavy}}

\vspace{.3cm}

\begin{tabular}{|c|c|c|} \hline
           & $\bar{m}(\bar{m})$ & $\bar{m}(M_Z)$ \\ \hline
S. Narison & $\bar{m}_b(\bar{m}_b) = 4.29 \pm 0.04$
           & $\bar{m}_b(M_Z) = 2.97 \pm 0.13$    \\
           & $\bar{m}_c(\bar{m}_c) = 1.28 \pm 0.04$
           & $\bar{m}_c(M_Z) = 0.52 \pm 0.09$    \\ \hline

S. Narison & $\bar{m}_b(\bar{m}_b) = 4.35 \pm 0.05$
           & $\bar{m}_b(M_Z) = 3.03 \pm 0.13$    \\
           & $\bar{m}_c(\bar{m}_c) = 1.31 \pm 0.06$
           & $\bar{m}_c(M_Z) = 0.54 \pm 0.10$    \\ \hline

S. Titard     & $\bar{m}_b(\bar{m}_b) = 4.397^{+0.018}_{-0.033}$
              & $\bar{m}_b(M_Z) = 3.07 \pm 0.11$    \\
F.J. Yndur\'ain & $\bar{m}_c(\bar{m}_c) = 1.306^{+0.022}_{-0.035}$
              & $\bar{m}_c(M_Z) = 0.52 \pm 0.08$    \\ \hline

M. Neubert & $\bar{m}_b(\bar{m}_b) = 4.37 \pm 0.09$
           & $\bar{m}_b(M_Z) = 3.04 \pm 0.17$    \\
           & $\bar{m}_c(\bar{m}_c) = 1.17 \pm 0.12$
           & $\bar{m}_c(M_Z) = 0.45 \pm 0.14$    \\ \hline \hline

Crisafulli et al.~\cite{CRI} & $\bar{m}_b(\bar{m}_b) = 4.17 \pm 0.06$
                     & $\bar{m}_b(M_Z) = 2.93 \pm 0.08$    \\  \hline

Crisafulli et al.~\cite{VG} & $\bar{m}_b(\bar{m}_b) = 4.15 \pm 0.07$
                     & $\bar{m}_b(M_Z) = 2.91 \pm 0.09$    \\  \hline

El-Khadra et al. & $\bar{m}_c(\bar{m}_c) = 1.36 \pm 0.19$
                 & $\bar{m}_c(M_Z) = 0.61 \pm 0.15$    \\  \hline

Davies et al. & $\bar{m}_b(\bar{m}_b) = 4.13 \pm 0.11$
              & $\bar{m}_b(M_Z) = 2.89 \pm 0.12$  \\ \hline \hline

{\it mean} & $\bar{m}_b(\bar{m}_b) = 4.33 \pm 0.06$
           & $\bar{m}_b(M_Z) = 3.00 \pm 0.12$    \\
           & $\bar{m}_c(\bar{m}_c) = 1.30 \pm 0.08$
           & $\bar{m}_c(M_Z) = 0.52 \pm 0.10$    \\ \hline
\end{tabular}
\end{center}

\end{table}


     I have summarized in tables \ref{heavytop}, \ref{heavy} and \ref{light}
the result for the running of the quark masses to $M_Z$. I mean by NLO
connection between the perturbative pole mass and the running mass
dropping the $O(\as^2)$ term, running to two loops and matching at one loop,
i.e. strong gauge coupling and masses continuous at
$\mu_{th} = \bar{m}(\bar{m})$. Three loops running and matching as expressed
in \Eq{continuous} with $x=0$ correspond to NNLO. For consistency with the
original works we can only do the
running for the bottom and the charm quarks just to NLO. For the
light quarks the running is consistent to NNLO using the threshold masses,
$\bar{m}(\bar{m})$, of the bottom and charm quarks determined at NLO.
I propagate the errors in the running in such a way we maximize them.

     It is informative to notice that the running mass of the top quark
is shifted about $7(GeV)$ down from its perturbative pole mass that is
of the order of its error. Therefore it is important
to clarify which mass CDF and D\O\ are talking about.
I have decoupled the top quark at $M_Z$ otherwise it makes no sense
to run the top down. This fact shifts down slightly the strong coupling
constant in $M_Z$, from $\as^{(5)}(M_Z) = 0.118 \pm 0.006$ we get
$\as^{(6)}(M_Z) = 0.117 \pm 0.006$ but has no effect on the masses
because the errors screen the difference between the theory with
5 and 6 flavours. Curiously the running of the top to $M_Z$ cancels
the difference between the running and the pole mass.

     One has to be very careful in comparing the running of the masses
obtained from QCD Sum Rules and those obtained from lattice
because I took different values for the strong coupling constant
at $M_Z$. In addition, we have to remember that the error in the
running is dominated by the error in the strong coupling constant.
However it is impressive to notice the good agreement
of the results obtained in lattice~\cite{DA,CRI}
with the running of the masses from QCD Sum Rules.
Even, the APE-Collaboration~\cite{CRI}, has improved recently its
result by increasing the statistics on the lattice~\cite{VG}.

     We can now play a game combining the light quark masses
of table \ref{light} with the ratios obtained from $\chi$PT. The mean
value of the strange quark mass together with the Bijnens et at.~\cite{BI}
result gives

\beq
   \frac{2 \bar{m}_s}{\bar{m}_u+\bar{m}_d} = 33. \pm 12.,
\eeq
      in agreement with the $\chi$PT result. Being conservative we
can also get for the up and down quarks
$\bar{m}_u(1GeV) = (3. \pm 2.) MeV$ and $\bar{m}_d(1GeV) = (9. \pm 2.) MeV$
that translate into $\bar{m}_u(M_Z) = (1.5 \pm 1.2) MeV$ and
$\bar{m}_d(M_Z) = (4.1 \pm 1.7) MeV$.


\begin{table}[hbtp]
\begin{center}

\caption{Running of the light quark masses to $M_Z$. For masses
extracted from QCD SR  $\as^{(5)}(M_Z) = 0.118 \pm 0.006$,
for lattice $\as^{(5)}(M_Z) = 0.115 \pm 0.002$. First box is NLO running,
second and third boxes are NNLO running.
\label{light}}

\vspace{.3cm}

\begin{tabular}{|c|c|c|} \hline
           & $\bar{m}(1GeV)$ & $\bar{m}(M_Z)$ \\ \hline
S. Narison & $\bar{m}_s(1GeV) = 222. \pm 22.$
           & $\bar{m}_s(M_Z) = 105. \pm 28.$    \\  \hline \hline
Allton et al. & $\bar{m}_s(1GeV) = 156. \pm 17.$
              & $\bar{m}_s(M_Z) = 78. \pm 15.$    \\  \hline
\end{tabular}

\vspace{.3cm}

\begin{tabular}{|c|c|c|} \hline
           & $\bar{m}(1GeV)$ & $\bar{m}(M_Z)$ \\ \hline
S. Narison & $\bar{m}_s(1GeV) = 197. \pm 29.$
           & $\bar{m}_s(M_Z) = 88. \pm 31.$    \\  \hline

Jamin / M\"unz & $\bar{m}_s(1GeV) = 189. \pm 32.$
               & $\bar{m}_s(M_Z) = 85. \pm 32.$    \\  \hline

Chetyrkin et al. & $\bar{m}_s(1GeV) = 171. \pm 15.$
                 & $\bar{m}_s(M_Z) = 75. \pm 23.$    \\  \hline \hline

{\it mean} & $\bar{m}_s(1GeV) = 186. \pm 30.$
           & $\bar{m}_s(M_Z) = 83. \pm 30.$    \\ \hline
\end{tabular}

\vspace{.3cm}

\begin{tabular}{|c|l|l|} \hline
Bijnens et al. & $(\bar{m}_u+\bar{m}_d)(1GeV) = 12.0. \pm 2.5$
               & $(\bar{m}_u+\bar{m}_d)(M_Z) = 5.4 \pm 2.2$    \\  \hline
Ioffe et al. & $(\bar{m}_d-\bar{m}_u)(1GeV) = 3. \pm 1.$
             & $(\bar{m}_d-\bar{m}_u)(M_Z) = 1.4 \pm 0.7$    \\  \hline
\end{tabular}

\end{center}

\end{table}


     To summarize, the running of the quark masses to the $M_Z$ energy
scale gives us a running top mass that is around its perturbative pole
mass, $\bar{m}_t(M_Z) = (180. \pm 13.)GeV$, for the bottom and
the charm quark we get $\bar{m}_b(M_Z) = (3.00 \pm 0.12) GeV$ and
$\bar{m}_c(M_Z) = (0.52 \pm 0.10) GeV$ respectively, while for the
strange quark we have a result affected by a big error
$\bar{m}_s(M_Z) = (83. \pm 30.)MeV$. The same happens for the
up and down quarks, we get $\bar{m}_u(M_Z) = (1.5 \pm 1.2) MeV$ and
$\bar{m}_d(M_Z) = (4.1 \pm 1.7) MeV$.

\section*{Acknowledgements}

I would like to acknowledge M.~Neubert, A.~Pich, K.~Hornbostel,
V.~Gim\'enez and F.~Botella for very useful discussions and
A.~Santamaria for carefully reading the manuscript.
This work was supported by the Conselleria d'Educaci\'o
i Ci\`encia de la Generalitat Valenciana,
CICYT (Spain) under grant AEN93-0234
and I.V.E.I. under grant 03-007.

\end{document}